# Suppression of Thermal Conductivity via Singlet-Dominated Scattering in $TmFeO_3$

M. L. McLanahan, D. Lederman, and A. P. Ramirez

*Physics Department, University of California Santa Cruz, Santa Cruz, CA 95064*

ABSTRACT

We measured the thermal conductivity of the rare-earth orthoferrites, $RFeO_3$, where $R$ = Eu, Gd, Tb, Dy, Ho, Er, Tm, and Yb from 3 K to 300 K and see an anomalous strong suppression for $TmFeO_3$ over most of the temperature range. Using a Debye thermal transport model, we demonstrate that this suppression is due to resonant scattering between phonons and the $Tm^{3+}$ $4f$ singlet crystal field levels. The implications of these results are discussed in context of thermal conductivity studies in quantum magnets.

Experimentally determining a condensed matter system's energy spectrum often involves interpreting thermodynamic data using a theoretical model of the system's thermal excitations. It is usually the case that such models are designed to render the excitations that can be measured, but in magnetic systems there exist excitations not observable using magnetic probes, namely spin singlets. While such excitations are rarely discussed for conventional magnets, singlets have been shown to dominate the low energy spectrum of the geometrically frustrated (GF) materials $SrCr_9Ga_4O_{19}$ [1] and $NiGa_2S_4$ [2], and they have been clearly identified in the magnetization plateau region of CeSb [3], and thus may be more determinative for the ground state of frustrated systems than has been previously considered.

A topic that illustrates the importance of understanding singlets is the purported spinon contribution to thermal conductivity of GF magnets. These systems exhibit no long-range-order even far below their Weiss temperatures and generally show an inelastic continuum of excitations in neutron scattering, a scenario consistent with spinon modes of a long-sought and exotic quantum spin liquid (QSL) ground state [4]. To more firmly establish the existence of spinon modes, however, measurements complementary to neutrons are needed. In ordered magnets, the specific heat ($C(T)$) is a useful probe of low-energy spin wave excitations, but in systems like GF magnets, where the degrees of freedom might include spinons, singlets, *and* localized modes due to quenched disorder (in addition to phonons) separation of the individual contributions can be challenging. Thus, for spinon detection, thermal conductivity ($\kappa(T)$) has been proposed as an alternative probe [4] because it has positive contributions only from mobile excitations, evading localized modes often seen in $C(T)$. Even in $\kappa(T)$, however, one still faces the problem of contributions to $\kappa(T)$ from singlets, which may be mobile. In addition, even though localized singlets cannot *carry* heat, they can *scatter* heat-carrying excitations, contributing *negatively* to $\kappa(T)$. Thus, a systematic understanding of the role of singlets in magnetic systems is essential for further progress in low-energy quantum magnetism.

Here we address explicitly the role of singlet excitations in thermal conductivity measurements. For GF magnets, we know neither the density nor the magnetoelastic coefficients so, to do a controlled study, we consider the singlets created by valence electrons on an ion. Among systems possessing well-defined localized singlet states are rare-earth ions, whose 4*f*-derived crystalline electric field (CEF) levels dictate thermodynamic behavior and thus should play an important role in $\kappa$-measurements. These levels, which can be either magnetic or non-magnetic, have been discussed as resonantly scattering thermal phonons, thus leading to strong suppression of $\kappa$, e.g. in $NdMnO_3$, $TbMnO_3$, $Tb_3Ga_5O_{12}$, and $Tb_2Ti_2O_7$ [5-7]. The interactions between the $R^{3+}$-$4f$ and $Fe^{3+}$-$3d$ electrons in the rare-earth orthoferrites $R$FeO$_3$ [8, 9], where $R$ is a rare-earth ion, form a family rich in magnetic phenomena, most notably a $\pi/2$ spin rotation of the $Fe^{3+}$ moment at temperatures well below the Fe-spin Néel temperature. Because the $R$FeO$_3$ series can be made with every $R$-member, they present a variety of different CEF level schemes and some, like $HoFeO_3$, show contributions from resonant scattering between phonons and CEF levels [10]. Of particular interest is $TmFeO_3$, the focus of this work, where the $Tm^{3+}$ ground multiplet splits into a series of low energy singlets, a unique level scheme among $R$FeO$_3$ compounds [11]. This feature presents an opportunity to study the effects of phonon-singlet scattering in a system where other magnetic contributions are minimized.

In this letter we present a thermal conductivity study of the $R$FeO$_3$ family where $R$ = Eu, Gd, Tb, Dy, Ho, Er, Tm, and Yb. In particular, we observe a 'double peak' feature in $\kappa(T)$ of $TmFeO_3$ resulting from a strong suppression of the scattering length relative to the other members of the series. Using a simplified phonon conduction framework, we model $\kappa(T)$ for multiple $R$FeO$_3$ compounds considering only boundary, point-defect, phonon-phonon, and resonance scattering by CEF levels. Furthermore, we demonstrate that the suppression seen in $TmFeO_3$ thermal conductivity data can be explained by, and likely originates in, phonon scattering from the multiple low energy singlets of the $Tm^{3+}$ ground multiplet split by the CEF.

The $R$FeO$_3$ compounds crystallize in an orthorhombic structure belonging to the *Pbnm* space group. In the temperature region 600 K – 700 K, the $Fe^{3+}$ ions order antiferromagnetically along the crystallographic $a$-axis with canted ferromagnetism along the $c$-axis, i.e., $\Gamma_4(G_x, F_z)$ spin configuration, where $F_i$ ($G_i$) with $i = [x, y, z]$ refers to ferromagnetic (G-type antiferromagnetic) Fe moment along $[a, b, c]$-axis, respectively [8, 12, 13]. The $R^{3+}$ ions can have a moment aligned either parallel or antiparallel to the $Fe^{3+}$ canted moment due to the Fe-exchange field. As $R$FeO$_3$ compounds are cooled, the Fe spins may undergo a spin rotation i.e., $\Gamma_4(G_x, F_z) \Rightarrow \Gamma_2(F_x, G_z)$ for Pr [9, 14], Nd [14-16], Sm [17], Tb [18], Ho [19, 20], Er [19], Tm [21, 22], Yb [19], or $\Gamma_4(G_x, F_z) \Rightarrow \Gamma_1(0, G_z)$ for Ce [23] and Dy [24]. The spin rotation is not observed when $R$ is nonmagnetic e.g., $La^{3+}$, $Eu^{3+}$, $Lu^{3+}$. Due to $R$-$R$ interactions, $R^{3+}$ ions order in several $R$FeO$_3$ compounds for T < 10 K.

The $RFeO_3$ ($R$ = Eu-Yb) single crystals used in this study were obtained from the UCSC sample archive [25] with a typical sample geometry of 5 mm x 1 mm x 0.5 mm. The flux method used to grow these crystals was discussed by Remeika and Kometani [26]. Dc magnetization measurements on our samples were carried out, confirming the expected distinct magnetic features of (Eu-Yb)$FeO_3$ compounds (see supplemental information [27], which includes references [8, 9, 11, 28, 29]). In addition, X-ray diffraction measurements were performed on $TmFeO_3$ to confirm crystal quality (see supplemental information [27]). Thermal conductivity was measured along the long axis using a steady-state method with a 2-thermometer-1-heater geometry. A 10-kΩ surface-mount resistor acted as the heater and bare element Cernox resistors were used as thermometers. The $\kappa(T)$ measurements as well as accompanying $C(T)$ measurements were performed in a Quantum Design PPMS. For $\kappa(T)$ measurements, we used our proprietary software but for $C(T)$ we used commercial software. Orientation of $\Delta T$ relative to the crystal axis does not significantly affect $\kappa$ in the absence of an external field. We tested for anisotropy by measuring $\kappa$ with $\Delta T$ both parallel and perpendicular to the $c$-axis in $TmFeO_3$ crystals and found minimal variation between the orientations (see supplementary information [27]). Previous measurements on $GdFeO_3$ [30] and $DyFeO_3$ [31] support this observation.

In Fig. 1 we show thermal conductivity of (Eu-Yb)$FeO_3$. At 300 K we find that $\kappa(T)$ is approximately 6-10 W/m·K, which is similar to other perovskites [5, 32]. As temperature decreases, $\kappa(T)$ increases until reaching a maximum and then subsequently decreases. This maximum occurs at $T \approx 15$ K for $R$ = Eu, Gd, Tb, Dy, Ho, Er, and $T \approx 20$ K for $R$ = Yb. Similar behavior is seen in the non-4$f$ isostructural analogue $YFeO_3$, which we have included in Fig. 1 with data taken from Zhao et al. [31]. The increase in $\kappa(T)$ is accompanied by a 'shoulder' occurring between 150 K to 90 K most notably for $R$ = Ho, Er, and Yb, a feature often associated with resonant scattering of phonons. A striking deviation from this general behavior is seen in $TmFeO_3$, where $\kappa(T)$ decreases on cooling below 135 K until a minimum is reached at 25 K followed by a small peak of 5.5 W/m·K at 7 K. This anomalous suppression in $\kappa(T)$ more closely resembles that of $TbMnO_3$, rather than the glasslike attenuation seen in $Tb_2Ti_2O_7$ and $EuTiO_3$ [33], and hints to an origin in the scattering of thermal phonons via CEF levels. We further discuss the strong suppression of $\kappa(T)$ observed in $TmFeO_3$ later.

The observed increase in $\kappa(T)$ upon cooling systems with $R \neq$ Tm is, like in most crystalline materials, due to the increase of phonon mean-free path resulting from reduced phonon-phonon scattering as thermal fluctuations decrease, with the eventual peak in $\kappa(T)$ indicating a crossover into boundary/defect-limited phonon transport where $\kappa(T)$ is expected to become proportional to $C(T)$. In $TmFeO_3$, however, the increase in $\kappa(T)$ on cooling below 25 K, is likely due to a different source of reduced phonon scattering, which is the focus of this work. In Fig. 2 we plot low temperature data for $C(T)$ alongside $\kappa(T)$. Rare-earth ordering is seen in $C(T)$ for $TbFeO_3$ and $ErFeO_3$ at 3.6 K and 4.5 K, respectively, but is absent in $\kappa(T)$. In contrast, the spin reorientation region in $YbFeO_3$ (6 K < T < 8 K) is observable in both $C(T)$ and $\kappa(T)$. This is notable because we do not observe signatures of spin reorientation in $\kappa(T)$ for the other $R$FeO$_3$ members. The spin reorientation transitions at temperatures above the phonon peak ($DyFeO_3$, $HoFeO_3$, $ErFeO_3$, and $TmFeO_3$) occur in a region where $\kappa(T)$ varies exponentially with $T$, and even though the spin reorientation for $TbFeO_3$ occurs at similar temperatures to $YbFeO_3$, $\kappa(T)$ is a factor of 6 larger. This suggests that the spin reorientation has a small effect on $\kappa(T)$, and is usually masked by the large changes in $\kappa(T)$. To investigate further we examine the thermal diffusivity, $D_T = \kappa/C$, in the low-temperature region.

In the simplest of terms, the thermal conductivity can be described by $\kappa = C v_p l/3$, where $C$, $v_p$, $l$ are the heat capacity per unit volume, mean phonon velocity, and the phonon mean-free path, respectively. As the phonon velocity is not expected to vary significantly with temperature, the ratio $\kappa(T)/C(T)$, i.e. $D_T$, is a convenient measure of an effective thermal transport length scale. In the limit of boundary-dominated scattering we expect $l$, and thus $D_T$, to be constant. This interpretation, however, must be used with care for systems that may have localized contributions to the specific heat. We plot the ratio $\kappa(T)/C(T)$ in Fig. 2. The general trend observed among the $R$FeO$_3$ compounds for $R \neq$ Tm is that $\kappa(T)/C(T)$ decreases monotonically with temperature. Contrary to these observations, $TmFeO_3$ exhibits the opposite behavior as $\kappa(T)/C(T)$ increases with decreasing temperature in the temperature range shown. A possible explanation for this is that $Tm^{3+}$ has no excited CEF levels below 15 K, while the behavior observed in the other compounds is due to localized contributions to the specific heat. In both scenarios, $\kappa(T)/C(T)$ continues to vary for temperatures below the phonon peak, suggesting that boundary scattering is not the dominant $\kappa$-limiting mechanism, and therefore not the cause of the depression in $\kappa(T)$ in $TmFeO_3$. To gain further insight into possible sources of phonon scattering observed, we analyzed $\kappa(T)$ within a scattering time framework.

The phonon scattering time, $\tau$, often depends on the phonon energy, and thus angular frequency $\omega$, with scaling dependent on the scattering source. As the mean-free path is proportional to the average scattering time, i.e. $l = v_p\tau$, it is more accurate to consider the frequency dependence of $l$ and sum over all phonon modes when calculating $\kappa(T)$. In the Debye approximation, the simple thermal conductivity expression used earlier can be written as,

$$\kappa(T) = \frac{v_p}{3}\int_0^{\omega_{\max}} C(\omega,T)l(\omega,T)\,d\omega, \tag{1}$$

where $\omega_{\max}$ is the phonon cutoff frequency. For comparison to experimental results, it is often useful to rewrite the above equation in terms of $\tau(\omega,T)$ as such,

$$\kappa(T) = \frac{k_B^4 T^3}{2\pi^2\hbar^3 v_p}\int_0^{\Theta_D/T} \frac{x^4 e^x}{(e^x-1)^2}\tau(x,T)\,dx, \tag{2}$$

where $k_B$ is the Boltzmann constant, $x = \hbar\omega/k_BT$ is a dimensionless variable, $\Theta_D = \hbar\omega_{\max}/k_B$ is the Debye temperature, and $C(\omega,T)d\omega$ is expressed in terms of the Debye model [34]. The average phonon scattering time is determined using Matthiessen's rule, i.e.

$$\tau^{-1} = \frac{v_p}{d} + \alpha\left(\frac{k_BT}{\hbar}\right)^4 x^4 + \beta\left(\frac{k_B}{\hbar}\right)^2 T^3x^2e^{-\Theta_D/bT} + \tau_{\text{res}}^{-1}\,, \tag{3}$$

where the terms we consider describe boundary, point-defect, phonon-phonon Umklapp, and resonance scattering events, respectively. Here, $\alpha$ and $\beta$ are proportionality constants, $b$ is a scaling constant of order unity, and $d$ is a length scale corresponding to the maximum phonon mean-free path in the crystal. Resonant scattering can be expressed as

$$\tau_{\text{res}}^{-1} = \sum_i \frac{\gamma_i T^4 x^4}{(\Delta_i^2 - T^2x^2)^2} F_i(T), \tag{4}$$

with $\gamma_i$ representing the CEF-level-phonon coupling strength and $\Delta_i k_B$ the energy difference of two CEF levels involved in the $i^{\text{th}}$ phonon-induced transition [35, 36]. Here, $F_i(T)$ represents the relative population of the CEF levels. While different variations of $F_i(T)$ have been used to analyze $\kappa(T)$ in a multitude of systems [37-41], we used the physically motivated form of $F_i(T)$ as equaling the difference in the fractional electron population of the scattering levels. This form has been used to successfully model $\kappa(T)$ in multilevel systems, including the rare-earth garnets [35, 42, 43]. Electron populations were calculated for the various compounds using Boltzmann statistics such that

$$F(T) = \frac{\exp(E_m/k_BT) - \exp(E_n/k_BT)}{Z(T)} \tag{5}$$

for a transition involving the $m^{\text{th}}$ and $n^{\text{th}}$ CEF levels, which the notation $(E_m, E_n)$ is used to denote this transition moving forward. Here, $Z(T)$ is the partition function composed of all considered CEF levels, where the CEF levels are obtained from optical and inelastic neutron scattering (INS) measurements [11, 44-48]. The large parameter space introduced by the inclusion of the different scattering terms can make the fitting process susceptible to overparameterization, thus reducing the confidence level. We reduce this uncertainty by first fitting a family of thermal conductivity curves to determine the fit parameters that should be similar among the different $R$FeO$_3$ crystals. Enforcing these estimates when fitting TmFeO$_3$ allows us to determine the origin of the different $\kappa(T)$ behavior without being influenced by large parameter discrepancies. To further improve fit confidence, we also reduce the number of fitting parameters by fixing $\Theta_D$ and $v_p$. The different $\Theta_D$ used in the fittings were obtained from work by Parida et al. [49], and a value of 4267 m/s was used for $v_p$ based on the average sound velocity in TmFeO$_3$ [50, 51]. Furthermore, due to the large energy scale of $R^{3+}$ 4$f$ electronic states, we consider only resonance scattering from the first few CEF levels originating from the $R^{3+}$ ground multiplet. Nonlinear curve fitting was performed via chi-squared minimization using the trust-region-reflective algorithm, and verifying that the Jacobian had full column rank, ensuring parameter identifiability.

In Fig. 3 we fit $\kappa(T)$ for DyFeO$_3$, HoFeO$_3$, ErFeO$_3$, and TmFeO$_3$ to the thermal conductivity model above with fitting results summarized in Table 1. Because DyFeO$_3$ shows no clear signatures of resonance scattering in $\kappa(T)$, the model is simplified by removing $\tau_{\text{res}}^{-1}$. Using the fit protocol described above, we find good agreement with $d = (3.58 \pm 0.12) \times 10^{-4}$ m, $\alpha = (1.09 \pm 0.04) \times 10^{-43}$ s$^3$, $\beta = (7.28 \pm 0.21) \times 10^{-18}$ s/K, and $b = 6.60 \pm 0.08$, consistent with the parameters reported for YFeO$_3$, with the exception of $\beta$ [10]. The discrepancy in $\beta$ can be traced to the previous YFeO$_3$ study considering an Umklapp scattering rate with $\tau_U^{-1} \propto \omega^3$, whereas we use a more physical $\tau_U^{-1} \propto \omega^2$ dependence [52, 53]. Using the DyFeO$_3$ parameters as initial estimates, we then fit HoFeO$_3$ and ErFeO$_3$, for which resonance scattering must be considered.

In HoFeO$_3$, CEF calculations predict that the Ho$^{3+}$ $^5I_8$ ground multiplet splits into a series of quasi-doublets [46] with optical measurements reporting a first excited doublet of $E_2/k_B = 124.7$ K and a ground doublet splitting of $E_1/k_B = 8.9$ K [44, 45]. Furthermore, heat capacity measurements reveal a Schottky anomaly that can be fit considering 3-4 degenerate levels around 200 K [21] or 230 K [54] in addition to the ground and first excited doublets. To fit our thermal conductivity data, it was necessary to consider a transition from the first excited doublet to another excited state, with the reported 4-fold degenerate level at 230 K resulting in the best fit. The corresponding fit parameter governing the CEF-level-phonon coupling, $\gamma = 5.5 \times 10^{11}$ s$^{-1}$, is similar to those found for the rare-earth 4$f$ electrons in garnets, i.e. DyAlG [35].

The above techniques are then applied to $ErFeO_3$, with $Er^{3+}$ being a Kramers ion. The $Er^{3+}$ $^4I_{15/2}$ ground multiplet splits into 8 Kramers doublets with approximate first and second excited state energies of $E_1/k_B = 67$ K and $E_2/k_B = 155.5$ K, respectively [44, 47]. The best fit was achieved using two resonance scattering terms involving $(E_1, E_2)$ and $(E_0, E_2)$, where we used the reported infrared-determined values to fix the CEF level separation energies, $\Delta_1 = 88.5$ K and $\Delta_2 = 155.5$ K, and considered only energy levels up to the second excited doublet for $F(T)$. From our fit, we find $\gamma_1 = 8.4 \times 10^{10}$ s$^{-1}$ and $\gamma_2 = 6.4 \times 10^9$ s$^{-1}$. The variation in $\gamma$ between $HoFeO_3$ and $ErFeO_3$ may result from a combination of: $\gamma$'s dependence on 4$f$ electron wave functions and contributions from multiple scattering centers. We note that the other scaling coefficients agree well with those of $DyFeO_3$ and $HoFeO_3$.

Unlike other $R$FeO$_3$ with non-Kramers $R^{3+}$ (e.g., $Tb^{3+}$ and $Ho^{3+}$) ions, $Tm^{3+}$ is unique in that the orthorhombic crystal field splits its $^3H_6$ ground multiplet into a series of isolated singlets, as opposed to quasi-doublets [48, 55]. In addition, the first few CEF levels of these singlets are low in energy (< 200 K), offering an environment abundant in potential scattering centers to include in $\tau_{\text{res}}$. We consider the CEF levels $E_1/k_B = 25.2$ K, $E_2/k_B = 56.2$ K, $E_3/k_B = 100.8$ K [11]. Without placing restrictions on the fitting parameters, we find that considering the $(E_1, E_3)$ transition fits $\kappa(T)$ generally well. Including the $(E_0, E_2)$ transition further improves the low temperature side of the fit (see supplemental information [27]). However, in both scenarios, some parameters deviate from those of the other $R$FeO$_3$ fits. In particular, $b \lesssim 3$ in both scenarios and when considering only $(E_1, E_3)$, $d$ is an order of magnitude smaller than the other $R$FeO$_3$ crystals. Because these parameters describe the Umklapp scattering and maximum phonon mean-free path, we would expect these parameters to be similar in the $R$FeO$_3$ family. Using the parameters and uncertainties found for the other $R$FeO$_3$ crystals, we create fit bounds for all scattering parameters except $\gamma_i$, the amplitude for resonant CEF scattering. Enforcing these fit parameter limits, we find it is necessary to include at least three $\tau_{\text{res}}^{-1}$ terms to accurately fit $\kappa(T)$, with transitions $(E_1, E_2)$, $(E_2, E_3)$, and $(E_0, E_3)$ resulting in the best fit (see Fig. 3). Comparing the $\gamma_i$ parameters (see Table 1), the dominant resonant scattering term involves the $(E_0, E_3)$ transition, which has an energy separation of $\hbar\omega/k_B = 100.8$ K. By the dominant phonon approximation ($\hbar\omega \approx 3.8k_BT$), such an energy separation should result in a dip in $\kappa(T)$ at $T \approx 25$ K, which is where we observe the strongest suppression.

We note that $TbFeO_3$ and $YbFeO_3$ also possess a shoulder-like feature in $\kappa(T)$, however data regarding their ground multiplet is more limited. To the authors' knowledge, only the first excited doublet of $YbFeO_3$ (~232 K [56]) has been measured, which alone is insufficient to fit $\kappa(T)$ in our temperature window. For $TbFeO_3$, INS measurements reveal the lowest-lying CEF level to be at either ~174 K [57] or ~197 K [55] with higher levels at ~290 K and ~406 K [58]. The best $\kappa(T)$ fit was performed by fixing one CEF level at 197 K while letting the other vary, with the optimal fit corresponding to a fit value of 247 K (see supplementary information [27]). However, since one of the CEF levels was a free parameter, we did not include those results to the bounds enforced on the $TmFeO_3$ fit parameters. Future investigations into the CEF levels of these crystals may give additional insight into the CEF-level-phonon coupling strength.

The above analysis supports the notion that the resonant scattering signature in $\kappa(T)$ of many of the $R$FeO$_3$ compounds is due to phonons scattering via the low-energy $R^{3+}$ $4f$ CEF levels from the ground multiplet. From studying a family of $\kappa(T)$ curves we found estimates for scattering parameters that should be similar among $R$FeO$_3$ crystals. Applying these estimates reduced the uncertainty due to overparameterization in this large parameter space, allowing us to identify the unique $Tm^{3+}$ CEF environment of $TmFeO_3$ as the source of the suppression observed. We present this explicitly in Fig. 3 where we plot the $TmFeO_3$ fit line without the resonance scattering terms and it resembles the $\kappa(T)$ behavior seen in the other $R$FeO$_3$ samples.

Our results show decisively that a system of spin-singlet states with energies comparable to heat-carrying phonons can result in strong suppression of the thermal conductivity and demonstrates that CEF levels without spin fluctuations can scatter phonons as effectively as their counterparts, e.g. doublets. Such a result may find application in low-$\kappa$ material design, especially for thermoelectric engineering, where $\kappa$ plays a key role for the refrigeration figure of merit. Here, an important quantity for assessing the magnitude of the suppression of $\kappa(T)$ is the CEF-level-phonon parameter, $\gamma_i$, representing the coupling of local lattice strain to the CEF levels of the scattering centers. The full amount of suppression will depend not only on this coupling constant but also on the density of scattering centers and the number of CEF levels with thermally-accessible energy levels [59, 60]. Thulium, in the orthoferrite structure, has a series of well isolated singlet CEF levels below 300 K and thus is especially effective in suppressing phonon heat transport.

Perhaps the most important outcome of our work will be to motivate an understanding of $\kappa(T)$ in spin systems that do not transition to long-range order, specifically those systems in which spinon excitations are thought to be present, namely geometrically frustrated magnets. These systems typically exhibit a distinct low-energy scale [61] that is independent of field at thermal energies [1, 2], thus strongly implying an origin in spin singlets. Such "inter-atomic" singlet-based states will not be characterized by the $\gamma_i$ parameter used above but may possess significant lattice coupling due to the strain dependence of the superexchange interaction. Future work should pursue a microscopic model of these excitations, in particular to address whether they are mobile and carry heat or localized and absorb heat. Certainly, any analysis of low-temperature thermal conductivity in geometrically frustrated systems should consider the effect of resonant scattering from singlets.

In summary, thermal conductivity of $R$FeO$_3$ single crystals was measured for $R$ = Eu-Yb. A general behavior typical of insulating crystals was observed with many also exhibiting signatures of resonance scattering in $\kappa(T)$. A deviation from this behavior was seen in TmFeO$_3$, revealing a 'double peak' feature and reduction in $\kappa(T)$. A purely thermal phonon transport model including boundary scattering, point-defect, Umklapp, and resonant scattering was sufficient to fit our data. Furthermore, we demonstrate through a robust $\kappa(T)$ analysis that the many low-energy singlet states of the $Tm^{3+}$ CEF-split ground multiplet is the origin of the strong suppression in $\kappa$ observed in TmFeO$_3$.

Acknowledgements – We thank K. Behnia for his comments on this work. This research was supported by U.S. Department of Energy Office of Basic Energy Science, division of Condensed Matter Physics grant DE-SC0017862.

FIGURES and TABLES

| $R$FeO3 | $d$ ($10^{-4}$ m) | $\alpha$ ($10^{-43}$ s$^3$) | $\beta$ ($10^{-18}$ s/K) | $b$ | CEF (K) | $\gamma_{ij}$ ($10^{10}$ s$^{-1}$) |
|---|---|---|---|---|---|---|
| $DyFeO_3$ | 3.58 $\pm$ 0.12 | 1.09 $\pm$ 0.04 | 7.28 $\pm$ 0.21 | 6.60 $\pm$ 0.08 | | |
| $HoFeO_3$ | 0.99 $\pm$ 0.01 | 1.36 $\pm$ 0.02 | 4.45 $\pm$ 0.11 | 7.25 $\pm$ 0.05 | $E_2/k_B = 124.7$ $E_3/k_B = 230$ | 55.41 $\pm$ 3.46 |
| $ErFeO_3$ | 2.21 $\pm$ 0.07 | 1.43 $\pm$ 0.17 | 5.52 $\pm$ 0.20 | 6.21 $\pm$ 0.23 | $E_1/k_B = 67$ $E_2/k_B = 155.5$ | 8.42 $\pm$ 1.05 |
| | | | | | $E_0/k_B = 0$ $E_2/k_B = 155.5$ | 0.64 $\pm$ 0.45 |
| $TmFeO_3$ | 1.88 $\pm$ 0.11 | 1.12 $\pm$ 0.19 | 5.24 $\pm$ 0.17 | 5.98 $\pm$ 1.09 | $E_1/k_B = 25.2$ $E_2/k_B = 56.2$ | 4.26 $\pm$ 0.18 |
| | | | | | $E_2/k_B = 56.2$ $E_3/k_B = 100.8$ | 34.06 $\pm$ 5.59 |
| | | | | | $E_0/k_B = 0$ $E_3/k_B = 100.8$ | 41.94 $\pm$ 2.24 |

Table 1. Parameters with uncertainties from thermal conductivity fits. CEF column corresponds to the pair of CEF levels considered in $\tau_{\mathrm{res}}^{-1}$, with multiple CEF rows indicating additional pairs included per $R$FeO3 fit.

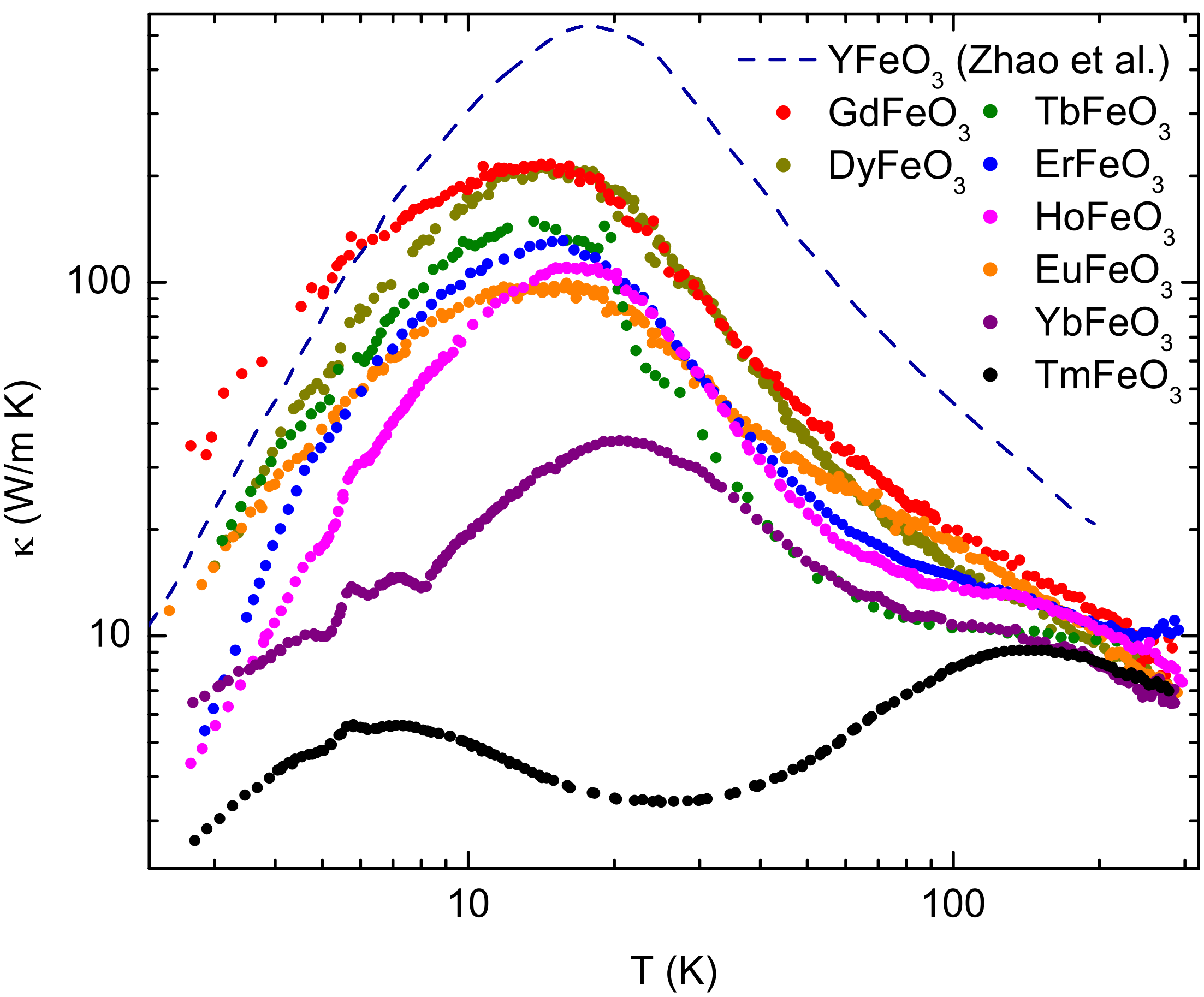

Figure 1. *R*$FeO_3$ thermal conductivity for *R* = Eu-Yb. $TmFeO_3$ exhibits a 'double peak' feature and is strongly suppressed compared to other *R*$FeO_3$ crystals. For completeness we include thermal conductivity data taken from Zhao et al. [31] for the non-4*f* isostructural analogue $YFeO_3$.

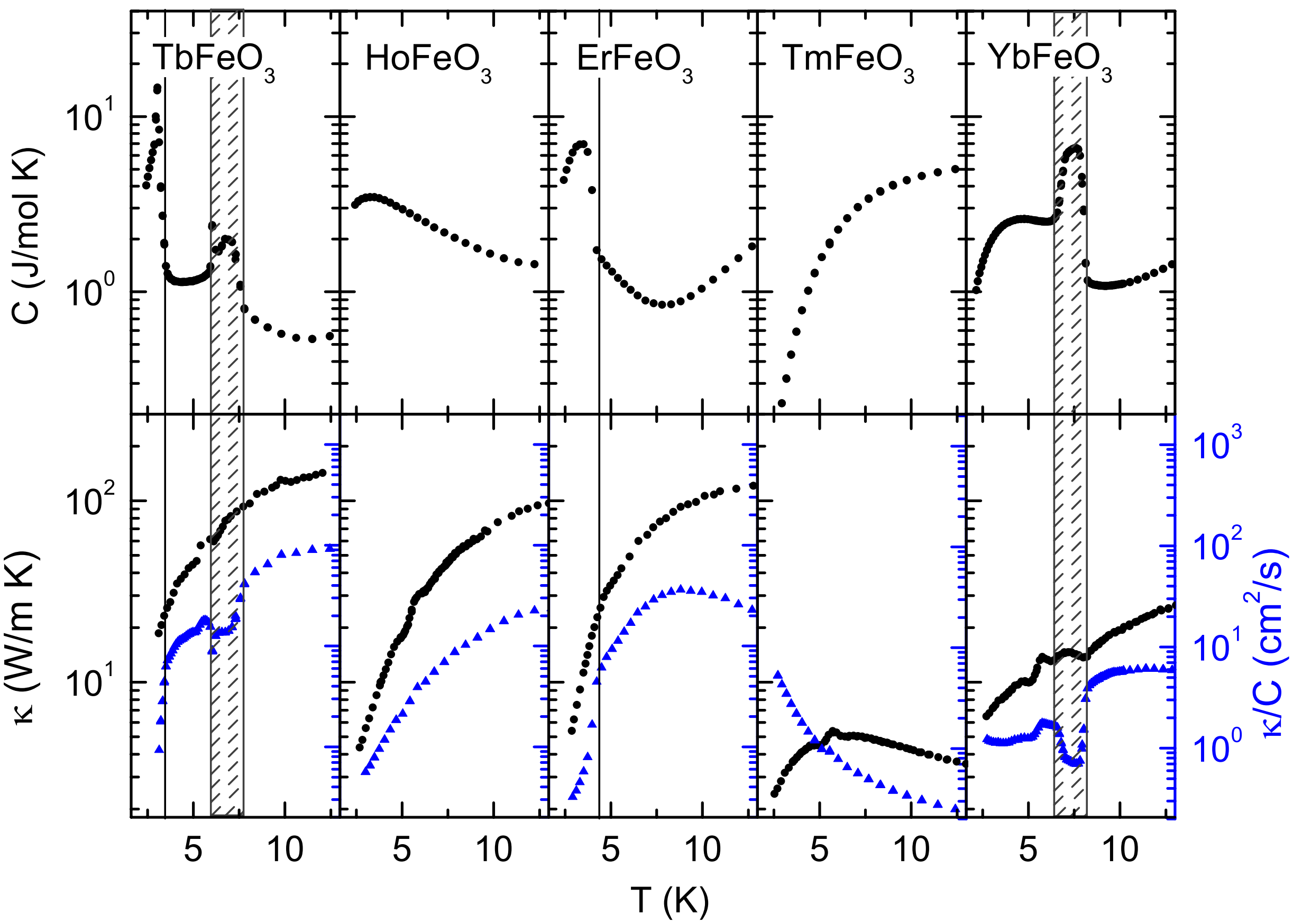

Figure 2. Low temperature heat capacity and thermal conductivity for select $R$FeO$_3$ compounds. Blue triangles correspond to the ratio $\kappa/C$. Vertical grey lines and dashed grey bars represent the $R^{3+}$ ordering temperature and $Fe^{3+}$ spin reorientation region, respectively. The other systems exhibit $Fe^{3+}$ spin reorientation at higher temperatures.

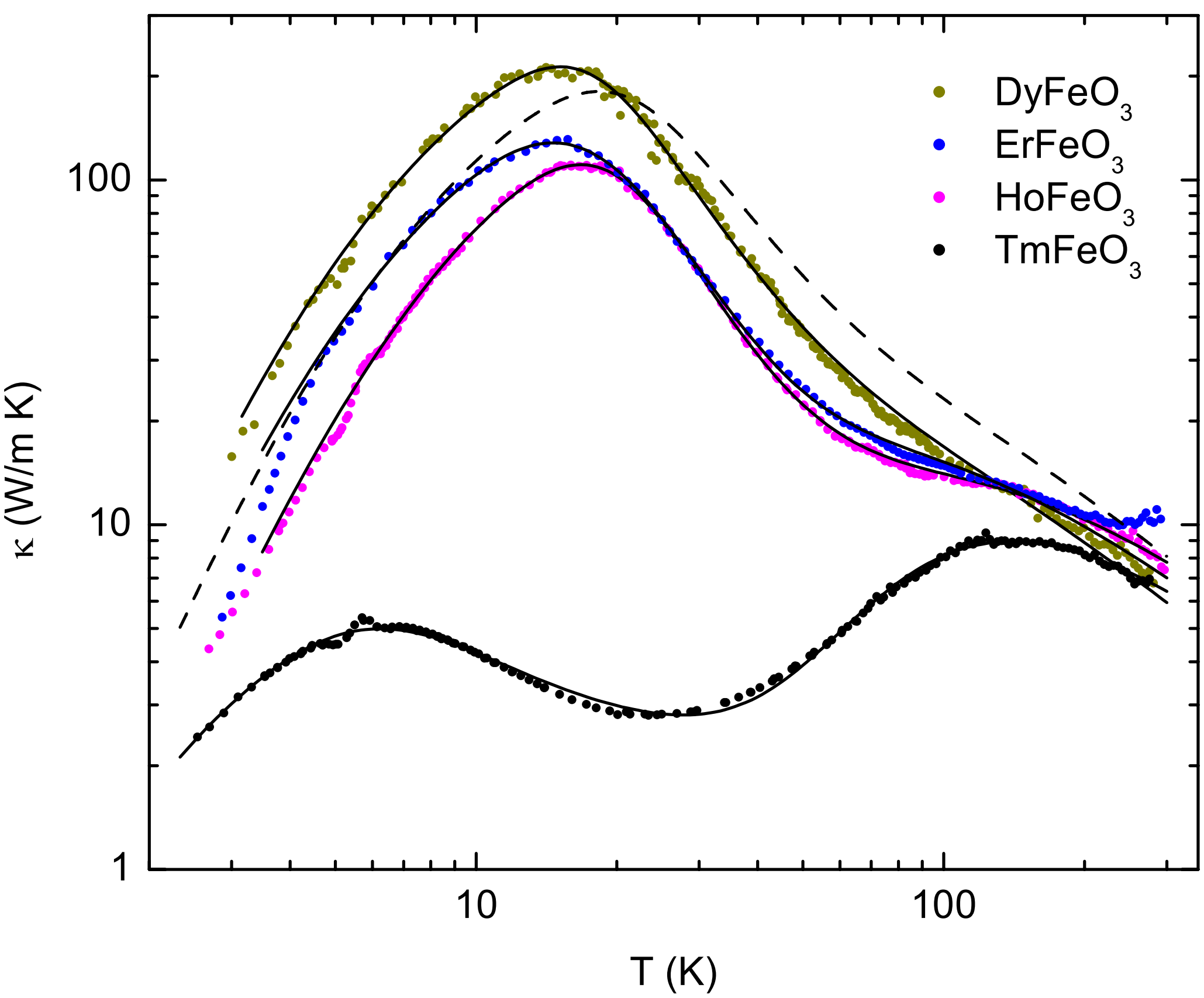

Figure 3. Thermal conductivity fits (solid black) using a phonon transport model including resonant scattering via CEF levels. The dashed line corresponds to our $TmFeO_3$ fit after removing the resonant scattering terms.

Supplementary Note: Suppression of Thermal Conductivity via Singlet-Dominated Scattering in $TmFeO_3$

M. L. McLanahan, D. Lederman, and A. P. Ramirez

*Physics Department, University of California Santa Cruz, Santa Cruz, CA 95064*

**Supplementary Note 1: Crystal Structure**

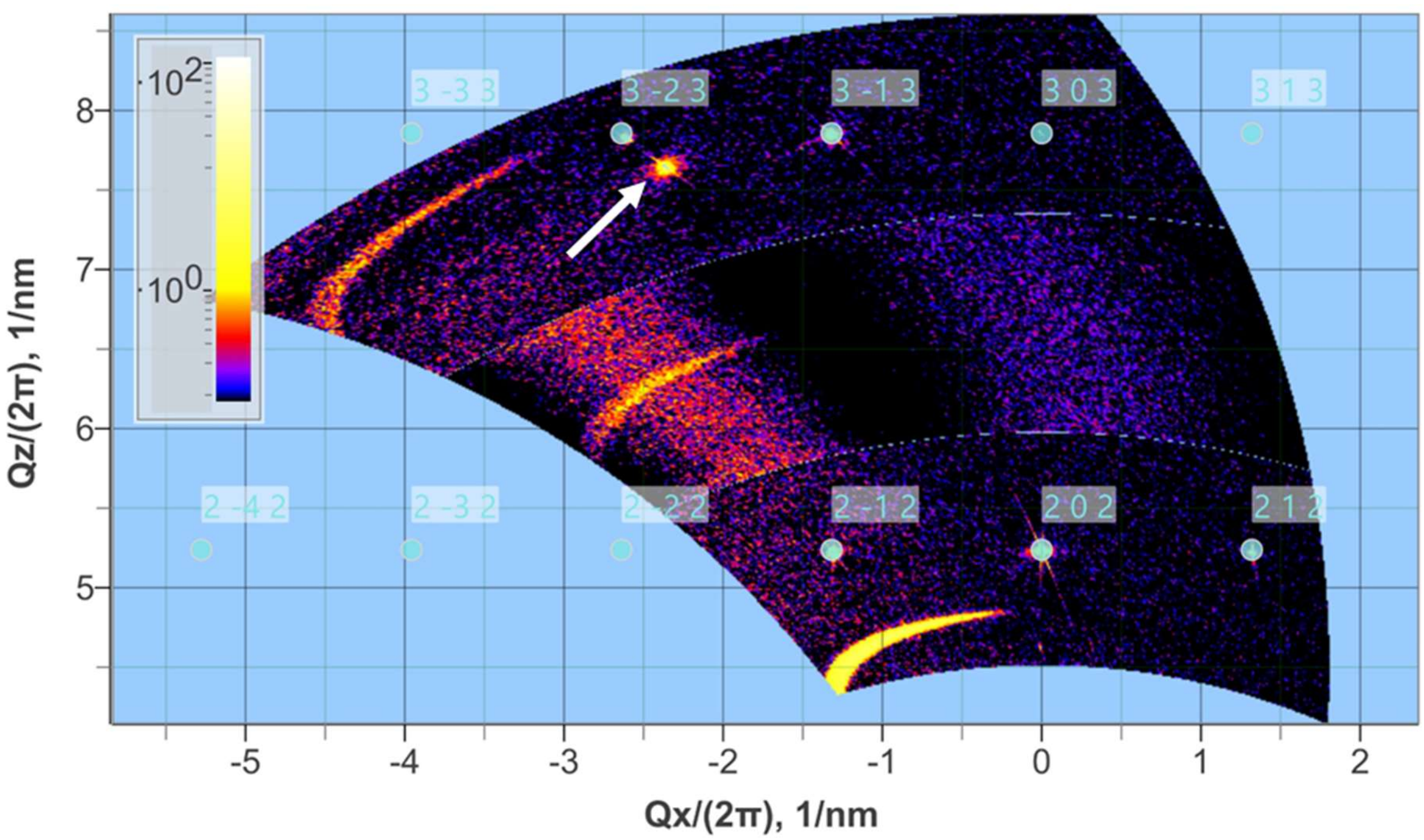

Figure S1. Reciprocal space map (RSM) of $TmFeO_3$ sample obtained using a 9 kW rotating anode diffractometer (Rigaku SmartLab II) at room temperature. The spots are labeled by the (hkl) Miller indices corresponding to the expected diffractions spots for $TmFeO_3$. The unlabeled bright spot indicated by the arrow corresponds to the $Al_2O_3$ single crystal used to mount the sample. This shows that the crystal is single phase, with the scattering planes oriented perpendicular to the (101) direction, and completely consistent with the crystalline structure measured in the literature [1, 2].

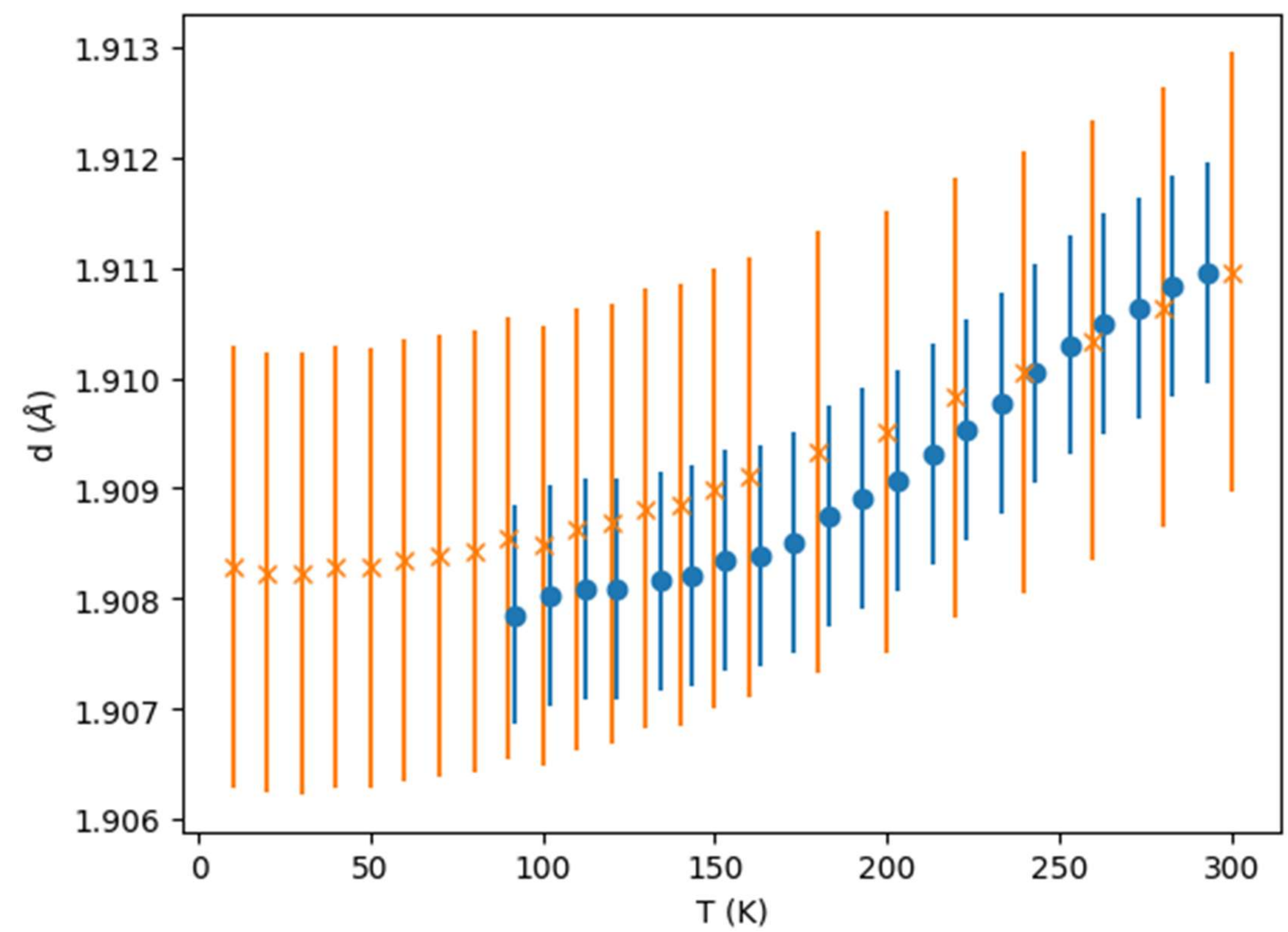

Figure S2. The (202) lattice constant of $TmFeO_3$ as a function of temperature measured at UCSC (solid circles) compared to the same lattice constant measured by Bombik et al. (x) [2]. Our data has been shifted by approximately -0.001 Å to take into account a small systematic alignment error. The two sets of data agree qualitatively within the uncertainties of the measurements.

**Supplementary Note 2: Dc Magnetization**

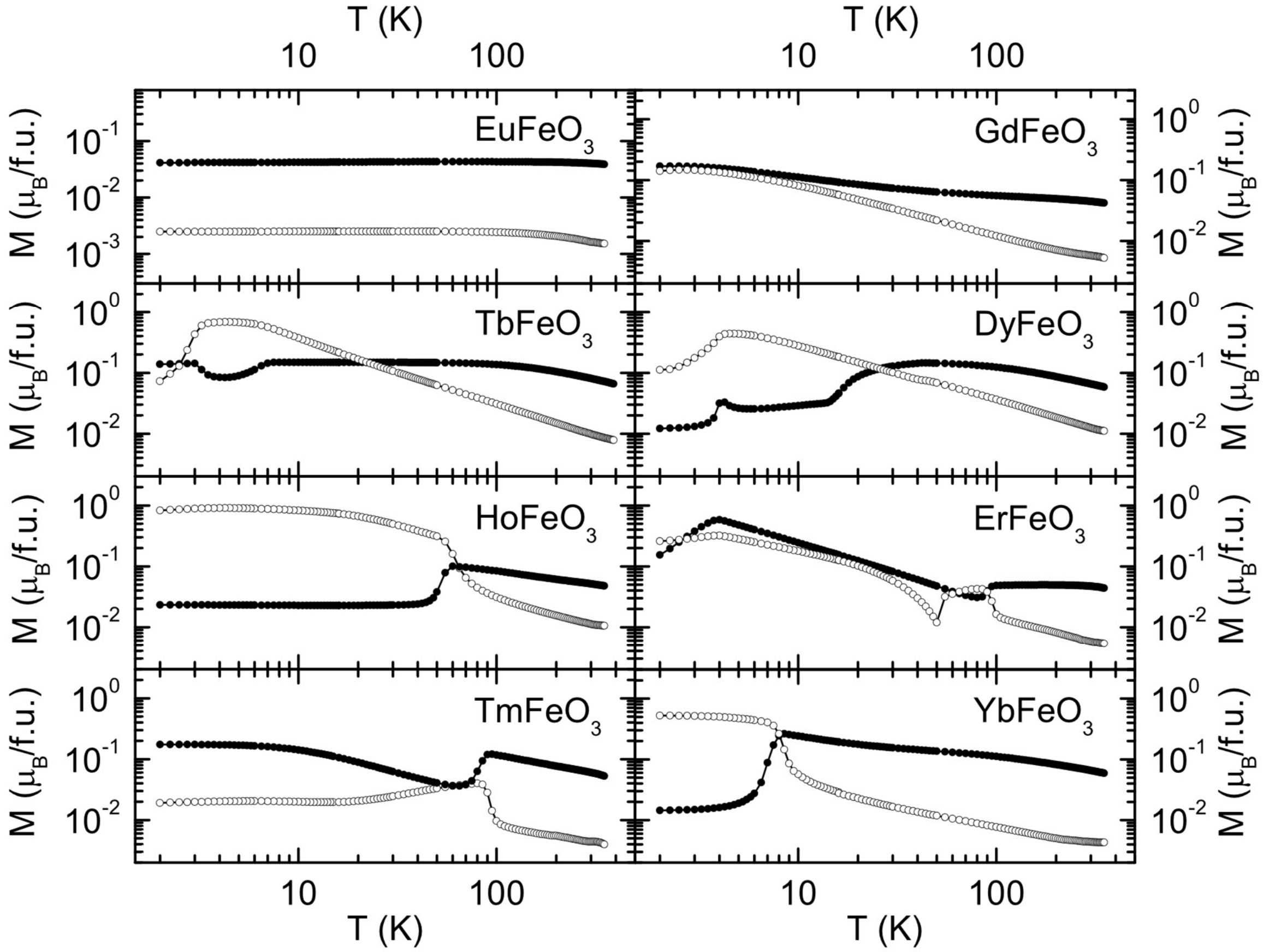

Figure S3. Dc magnetization of $R\mathrm{FeO_3}$, where $R$ = Eu -Yb, measured using a Quantum Design MPMS3. Measurements were performed with an external field of $H = 1$ kG for $H \parallel c$-axis (solid circles) and $H \perp c$-axis (open circles). The distinct magnetic features (e.g. $R^{3+}$ ordering, spin reorientation) observed in our $R\mathrm{FeO_3}$ compounds agree with those in the literature [3, 4].

## Supplementary Note 3: $TmFeO_3$ Thermal Conductivity

In Fig. S4 we plot $\kappa(T)$ for three different $TmFeO_3$ single crystals, Tm-1, Tm-2, and Tm-3. Both Tm-2 and Tm-3 came from the same crystal growth, while Tm-1 was from a different batch. All crystals had similar geometries of approximately 5.0 mm × 1.0 mm × 0.4 mm. Using a longitudinal flow method, the thermal conductivity was measured along the long axis of the crystal, which was perpendicular to the $c$-axis for both Tm-1 and Tm-2, and parallel to the $c$-axis for Tm-3. We find minimum variation in $\kappa(T)$ between all three samples, with largest differences of $< 1$ W/m·K, demonstrating reproducibility of the observed suppression in $\kappa(T)$ as well as the minimal effect of crystal orientation on $\kappa(T)$ in absence of a field. The Tm-3 data were presented in the manuscript.

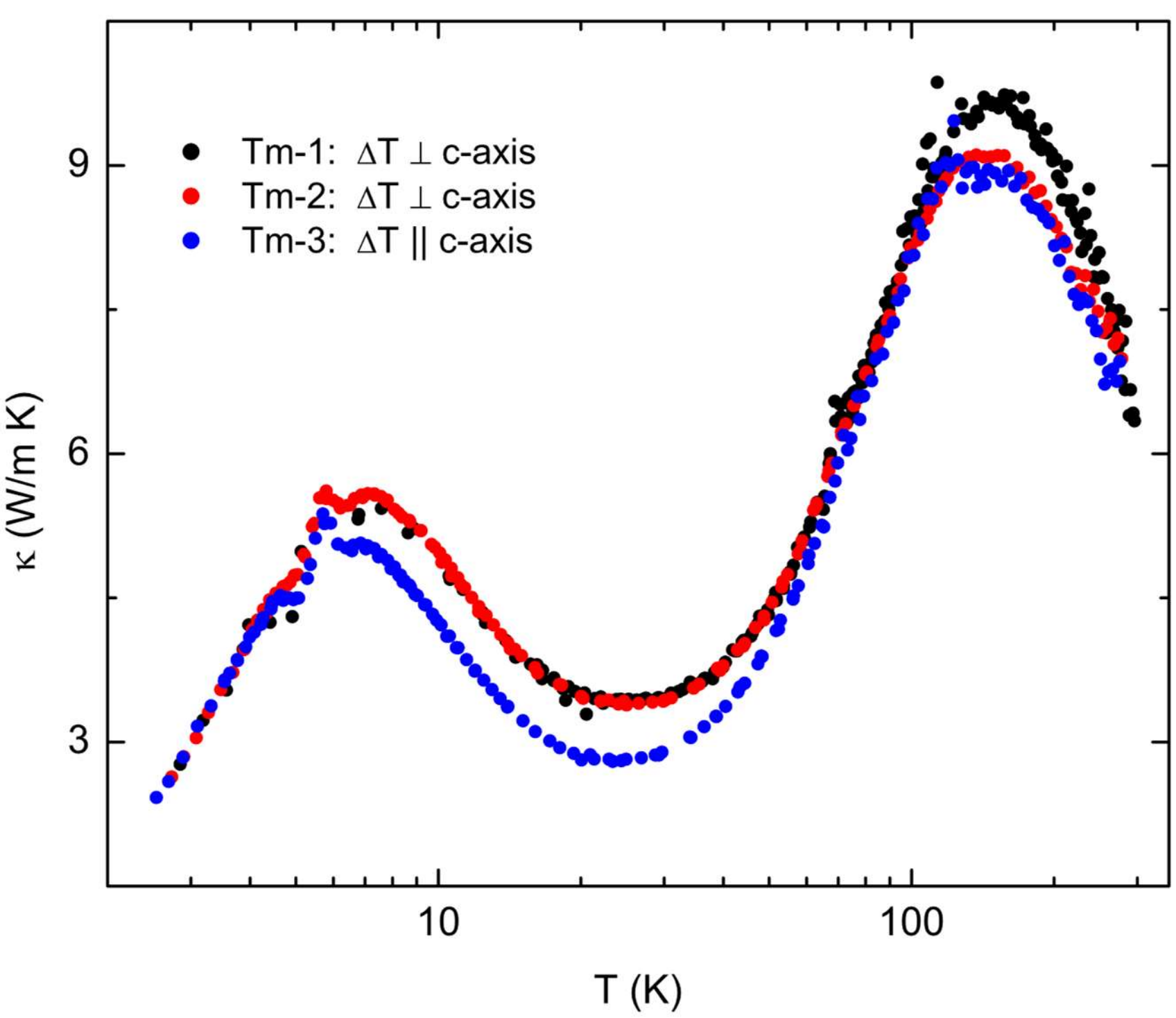

Figure S4. Thermal conductivity data for different $TmFeO_3$ crystals showing minimum variation among different samples and crystal orientation. Tm-2 and Tm-3 were from the same crystal growth batch.

The $TmFeO_3$ thermal conductivity data was fit to the Debye thermal transport equation (see manuscript), while varying the number of resonance terms included in $\tau_{\mathrm{res}}^{-1}$. The CEF energy levels were fixed using values reported by Malozemoff [5]. The best fits for scattering times including 1, 2, and 3 resonance terms are plotted in Fig. S5, with corresponding fit parameters listed in Table S1. Including 3 terms was necessary to sufficiently fit $\kappa(T)$ as well as have fitting parameters consistent with those from the thermal conductivity fits of the other $R FeO_3$ crystals.

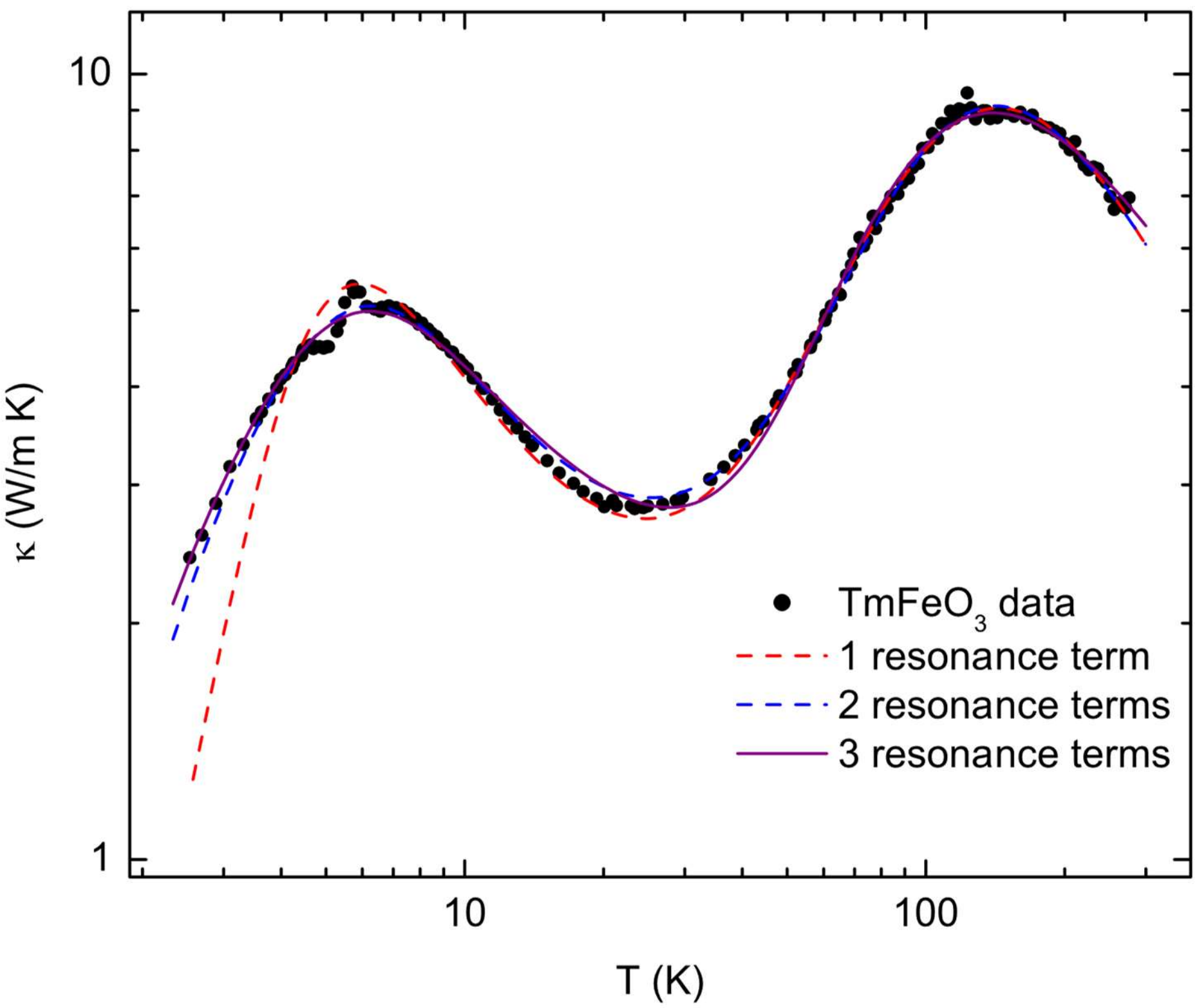

Figure S5. Thermal conductivity fits of $TmFeO_3$ sample Tm-3, where the number of resonance terms in the scattering time was varied. Three resonance terms were required to fit $\kappa(T)$ with parameters similar to the other $R FeO_3$ thermal conductivity fits.

| $TmFeO_3$ Fits | $d$ ($10^{-4}$ m) | $\alpha$ ($10^{-43}$ $s^3$) | $\beta$ ($10^{-18}$ s/K) | $b$ | CEF (K) | $\gamma_{ij}$ ($10^{10}$ $s^{-1}$) |
|---|---|---|---|---|---|---|
| 1 Res Term | 0.34 $\pm$ 0.01 | 2.95 $\pm$ 0.39 | 8.48 $\pm$ 1.01 | 1.91 $\pm$ 0.29 | $E_1/k_B = 25.2$ $E_3/k_B = 100.8$ | 141.52 $\pm$ 2.98 |
| 2 Res Terms | 1.35 $\pm$ 0.08 | 0.91 $\pm$ 0.15 | 7.16 $\pm$ 0.31 | 3.16 $\pm$ 0.33 | $E_1/k_B = 25.2$ $E_3/k_B = 100.8$ | 220.27 $\pm$ 4.46 |
| | | | | | $E_0/k_B = 0$ $E_2/k_B = 56.2$ | 2.65 $\pm$ 0.19 |
| 3 Res Terms | 1.88 $\pm$ 0.11 | 1.12 $\pm$ 0.19 | 5.24 $\pm$ 0.17 | 5.98 $\pm$ 1.09 | $E_1/k_B = 25.2$ $E_2/k_B = 56.2$ | 4.26 $\pm$ 0.18 |
| | | | | | $E_2/k_B = 56.2$ $E_3/k_B = 100.8$ | 34.06 $\pm$ 5.59 |
| | | | | | $E_0/k_B = 0$ $E_3/k_B = 100.8$ | 41.94 $\pm$ 2.24 |

Table S1. Parameters with uncertainties from $TmFeO_3$ thermal conductivity fits varying the number of resonant scattering terms. The CEF column corresponds to the pair of CEF levels considered in each resonant scattering term.

**Supplementary Note 4: $TbFeO_3$ Thermal Conductivity**

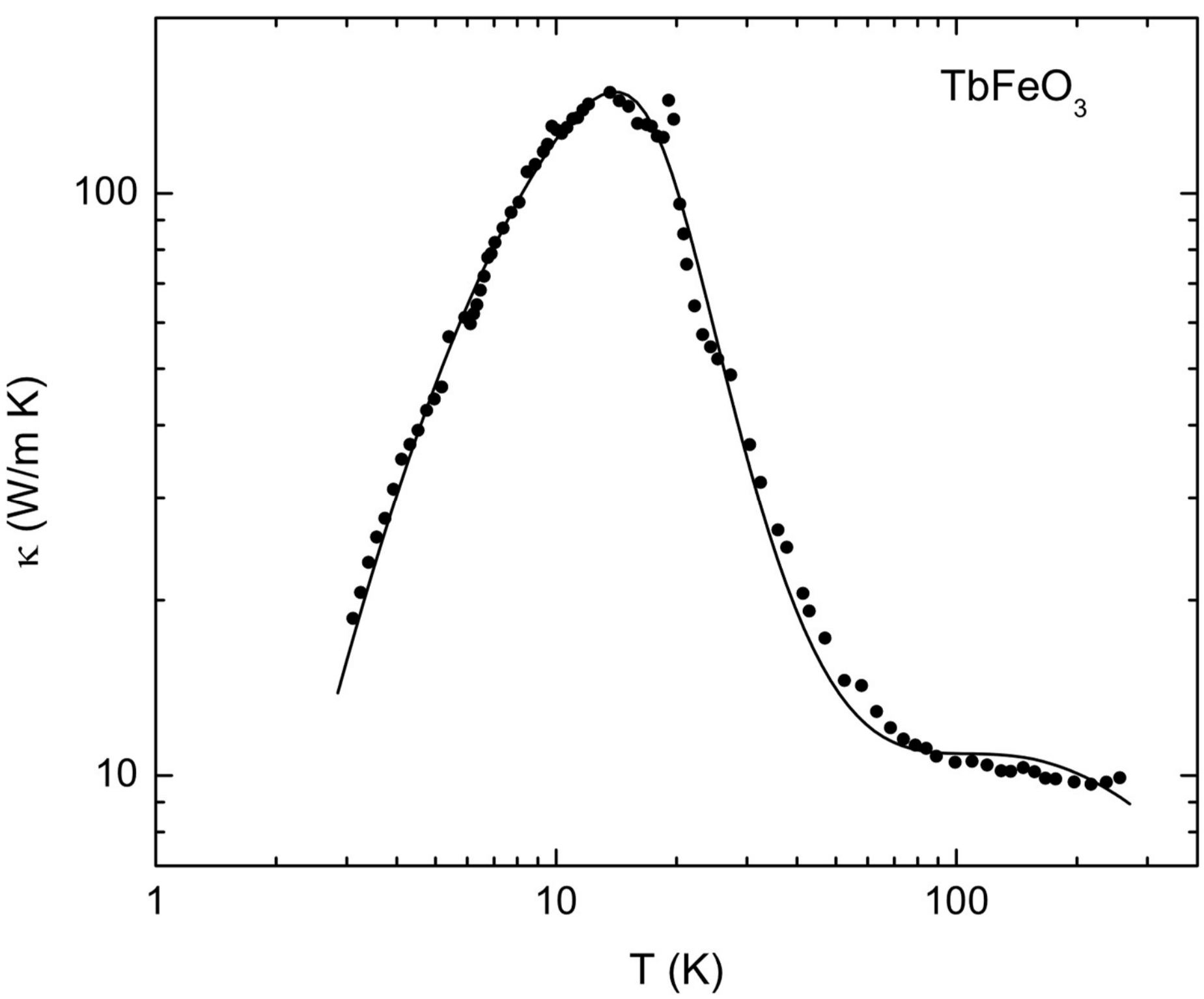

Figure S6. Thermal conductivity fit of $TbFeO_3$ considering one resonance term, where one CEF level was fixed at $E_m/k_B = 197.2$ K and the other a fitting parameter. Resulting fit parameters: $d = (3.24 \pm 0.09) \times 10^{-4}$ m, $\alpha = (1.74 \pm 0.00) \times 10^{-43}$ s$^3$, $\beta = (3.09 \pm 1.81) \times 10^{-18}$ s/K, $b = 8.28 \pm 1.39$, $\gamma_{mn} = (3.54 \pm 1.72) \times 10^{12}$ s$^{-1}$, and $E_n/k_B = (246.73 \pm 7.08)$ K.